\documentclass[12pt,osajnl2,preprint,showpacs,superscriptaddress,endfloats]{revtex4}  
\def\ListCaptions{y} 
\usepackage[draft]{hyperref} 
\usepackage{graphicx,amssymb,amsmath,amsfonts,bbm}

\newcommand{\bmsigma} {{\boldsymbol \sigma}}
\begin{document}
\title{Quantum characterization of bipartite Gaussian states} 
\author{D.~Buono}
\affiliation{Dipartimento di Scienze Fisiche Universit\`a ``Federico II'',
Monte Sant'Angelo, via Cintia,
I-80126 Napoli, Italy.} 
\author{G.~Nocerino}
\affiliation{Dipartimento di Scienze Fisiche Universit\`a ``Federico II'',
Monte Sant'Angelo, via Cintia,
I-80126 Napoli, Italy.} 
\author{V.~D'Auria}
\affiliation{Laboratoire Kastler Brossel, Ecole Normale Sup\'erieure,
Universit\'e Pierre et Marie Curie, CNRS, 4 place Jussieu, 75252 Paris, France.}
\author{A.~Porzio}
\email{alberto.porzio@na.infn.it}
\affiliation{CNISM UdR Napoli Universit\`a, Napoli, Italy.} 
\affiliation{CNR--SPIN, Monte Sant'Angelo, via Cintia,
I-80126 Napoli, Italy.}
\author{S.~Olivares}
\affiliation{CNISM UdR Milano Universit\`a, I-20133 Milano, Italy.} 
\affiliation{Dipartimento di Fisica dell'Universit\`a degli Studi di
Milano, I-20133 Milano, Italy.} 
\author{M.~G.~A.~Paris}
\affiliation{Dipartimento di Fisica dell'Universit\`a degli Studi di
Milano, I-20133 Milano, Italy.}
\affiliation{CNISM UdR Milano Universit\`a, I-20133 Milano, Italy.} 
\begin{abstract}
Gaussian bipartite states are basic tools for the realization of quantum
information protocols with continuous variables. Their complete characterization
is obtained by the reconstruction of the corresponding covariance matrix. Here
we describe in details and experimentally demonstrate a robust and reliable
method to fully characterize bipartite optical Gaussian states by means of a
single homodyne detector.  We have successfully applied our method to the
bipartite states generated by a sub-threshold type--II optical parametric
oscillator which produces a pair of thermal cross--polarized entangled CW
frequency degenerate beams.  The method provide a reliable reconstruction of the
covariance matrix and allows to retrieve all the physical information about the
state under investigation. These includes observable quantities, as energy and
squeezing, as well as non observable ones as purity, entropy and entanglement.
Our procedure also includes advanced tests for Gaussianity of the state and,
overall, represents a powerful tool to study bipartite Gaussian state from the
generation stage to the detection one.
\end{abstract}
\maketitle
\section{Introduction}
The quantum characterization of physical systems has a fundamental interest in its
own and represents a basic tool for the design of quantum protocols for
information processing in realistic conditions. In particular, the full
experimental reconstruction, at the quantum level, of optical systems opens the
way not only to high fidelity encoding/transmission/decoding of information, but
also to the faithful description of real communication channels and to precise
tests of the foundations of quantum mechanics \cite{LNP649,PO99,adv03,RMP09}.
\par
Among the systems of interest for quantum information processing we
focus on the class of bipartite optical states generated by parametric
processes in nonlinear crystals. These are Gaussian states and play a
crucial role in quantum information processing with continuous variables
\cite{EP03,BW03,AOP05,DDI06}.  Indeed, using single- and two-mode
Gaussian states, linear optical circuits and Gaussian operations, like
homodyne detection, several quantum information protocols have been
implemented, including teleportation, dense coding and quantum cloning
\cite{BVL05}.  In particular, Gaussian entangled states have been
successfully generated in the laboratories by type-II optical parametric
oscillators (OPO) below threshold \cite{dru90,zha99,kim92,prl99,tre05}.
In these OPO systems the parametric process underlying the dynamics is
well described, at least not too close to the threshold, by bilinear
Hamiltonian, thus the output states are Gaussian and they are completely
characterized by the first and second moments of their quadratures, i.e.
the covariance matrix.
\par
In this paper we address characterization of bipartite Gaussian states
and review in details a scheme to fully reconstruct the Gaussian output
from an OPO below threshold, which has been proposed in the recent years 
\cite{sh05,sh07} successfully implemented experimentally \cite{CM09}. In 
the present contibution we give a more accurate description of the experiment 
and data analisys and, in particular, we pay attention to advanced Gaussianity
test beyond the simple check of Kurtosis. 
Our method relies on a single homodyne detector: it provides the
full reconstruction of the covariance matrix (CM) by exploiting the possibility
of optically combining the two frequency degenerate OPO signal and idler beams
and then measuring suitable quadratures on the obtained auxiliary modes. Once
the CM is obtained one may retrieve all the quantities of interest on the state
under investigation, e.g. energy and squeezing, including those not
corresponding to any observable quantity like purity, entropy, entanglement,
and mutual information. Quantum properties are discussed in view
of the possible use of these states in quantum communication protocols. In particular,
we address the dependence of mutual information as a function of the bi--partite
system total energy.
Of course, a bipartite state is fully characterized by its covariance matrix if and
only if it is a truly Gaussian one. Usually one assumes that the state to be
processed has a Gaussian character because the interaction Hamiltonians are
approximated by bilinear ones and this is often an excellent approximation
\cite{pax99}. In turn, the resulting evolution corresponds a Gaussian
operations. On the other hand, it is known that nonGaussian dynamics may occur
when the OPO approaches the threshold \cite{opt05,K09} and when phase diffusion
\cite{Frz06,Hag09} is present during the propagation and/or the detection
stages. Therefore, in order to avoid any possible experimental issue
\cite{bow03,bow04,wen04,lau05}, a preliminary check on the Gaussian character of
the signal is crucial to ensure that the actual measured CM fully
characterizes the quantum state. For the first time, in this paper, CM data analisys
includes advanced statistical tests to assess Gaussianity \cite{opt05,jarda09}
of the state. 
The complete characterization strategy represents a
powerful tool to study bipartite Gaussians state from the generation stage 
to the detection one. 
\par
The paper is structured as follows. In section \ref{s:charGS} we
introduce the formalism used throughout the paper and, in particular, we
review two-mode Gaussian states and their covariance matrix as well as 
the relations among the CM elements and some physical quantities of 
interest, such as the purity, the entropy and the entanglement. The
method to reconstruct the CM is described in section \ref{s:rec}, while
section \ref{s:exp} is devoted to the details of our experimental
implementation. The analysis of the data and the results are discussed in 
details in Sections \ref{s:Gtest}, \ref{s:tomo} and \ref{s:results}.
In particular, test of Gaussianity are illustrated in Section
\ref{s:Gtest} and results from full quantum tomography in Section
\ref{s:tomo}. Section \ref{s:concl} closes the paper with 
some concluding remarks.
\section{Two-mode Gaussian states}
\label{s:charGS}
A $n$-mode state $\varrho$ of a bosonic system is Gaussian if its
characteristic function $\chi[\varrho](\boldsymbol{\lambda}) = \mathrm{Tr}[\varrho
D(\boldsymbol{\lambda})]$ has a Gaussian form, 
$D(\boldsymbol{\lambda}) = \bigotimes_{k=1}^n D_k(\lambda_k)$
being the $n$-mode displacement operator with
$\boldsymbol{\lambda} =
(\lambda_1,\ldots,\lambda_n)$, $\lambda_k \in \mathbb{C}$,
and $D_k(\lambda_k)=\exp \{\lambda_k a^\dag_k -
\lambda_k^* a_k \}$ denoting single-mode displacement operators
\cite{AOP05}.
Gaussian states are completely characterized by the first and second
statistical moments of the quadrature field operators, i.e. by the
vector of mean values and by the covariance matrix (CM). Since in this paper we 
focus on two-mode Gaussian states of the radiation field, in this section we
review the suitable formalism to describe the system. We also assume,
since this is the case in our experimental implementation, that the mean
values of quadratures are zero.
Upon introducing the vector of canonical operators $\boldsymbol{R}%
=(x_{a},y_{a},x_{b},y_{b})$, in terms of the mode operators
$\hat{a}_{k}$, $k=a,b$, $
\hat{x}_{k} =\frac{1}{\sqrt{2}}(\hat{a}_{k}^{\dag
}+\hat{a}_{k})$, $ 
\hat{y}_{k}=\frac{i}{\sqrt{2}}(\hat{a}_{k}^{\dag
}-\hat{a}_{k})$
the CM ${\boldsymbol{\sigma }%
}$ of a bipartite state is the real symmetric definite positive block matrix: 
\begin{equation}
\boldsymbol{\sigma }=\left( 
\begin{array}{c|c}
A & C \\ \hline
C^{T} & B%
\end{array}%
\right)  \label{CM}
\end{equation}%
with $\sigma _{hk}=\frac{1}{2}\langle \{R_{k},R_{h}\}\rangle -\langle
R_{k}\rangle \langle R_{h}\rangle $ being $\{f,g\}=fg+gf$. Matrices $A$, $B$
and $C$ are $2\times 2$ real matrices, representing respectively the
autocorrelation matrices of mode $a$ and $b$ and their mutual correlation
matrix. It can be observed that each block $A$, $B$ and $C$ can be written
as the sum of two matrices, one containing the product of mean values of $%
\left\langle R_{k}\right\rangle $ and the other contain the mean value of
products of operators $\langle \{R_{k},R_{h}\}\rangle $.  
\par 
Once the CM is known, all the properties of $\varrho $ may be described
and retrieved. As for example, the positivity of $\varrho$, besides
positivity of the CM itself, impose the constraint
\begin{equation}
\bmsigma +\frac{i}{2}\boldsymbol{\Omega \geq }0,
\label{Heisenberg principle}
\end{equation}%
where
$\boldsymbol{\Omega }=\boldsymbol{\omega }\oplus \boldsymbol{\omega }$
is the two-mode symplectic matrix, given in terms of
$\boldsymbol{\omega} \equiv\hbox{adiag}[1,-1]$.
Inequality (\ref{Heisenberg principle}) is equivalent to the Heisenberg
uncertainty principle and to positivity, and ensures that ${\boldsymbol{\sigma }}$ 
is a \textit{bona fide} CM. 
\par
A relevant result concerning the actual expression of a CM is that for
any two-mode CM ${\boldsymbol{\sigma }}$, there exists a  (Gaussian)
local symplectic operation $S=S_{1}\oplus S_{2}$
that brings ${\boldsymbol{\sigma }}$ in its standard form, namely 
\cite{sim00,dua00}: 
\begin{equation}
S^{T}{\boldsymbol{\sigma }}S=
\left( 
\begin{array}{c|c}
\tilde A & \tilde C \\ \hline
\tilde C^{T} & \tilde B%
\end{array}%
\right) ,
%
\end{equation}%
where $\tilde A = {\rm diag}[n,n]$, $\tilde B = {\rm diag}[m,m]$,
$\tilde C = {\rm diag}[c_{1},c_{2}]$,
with $n$, $m$, $c_{1}$ and $c_{2}$ determined by the four local
symplectic invariants $I_{1}\equiv \det (A)=n^{2}$, $I_{2}\equiv \det
(B)=m^{2}$, $I_{3}\equiv \det (C)=c_{1}c_{2}$, $I_{4}\equiv \det ({%
\boldsymbol{\sigma }})=\left( nm-c_{1}^{2}\right) \left( nm-c_{2}^{2}\right) 
$. If $n=m$, the matrix is called symmetric and represents a symmetric 
bipartite state where the energy is equally distributed between the two
modes.
\par
By using the symplectic invariants the uncertainty relation (\ref{Heisenberg
principle}) can be expressed as: 
\begin{equation}
I_{1}+I_{2}+2I_{3}\leq 4I_{4}+\frac{1}{4}.  \label{Heis rel. sympl invariant}
\end{equation}%
It is useful to introduce the symplectic eigenvalues, denoted by $d_{\pm }$ with 
$d_{-}\le d_{+}$, which in terms of symplectic invariants read as
follows \cite{ser:04}
\begin{equation}
d_{\pm }=\sqrt{\frac{\Delta ({\boldsymbol{\sigma }})\pm \sqrt{\Delta ({%
\boldsymbol{\sigma }})^{2}-4I_{4}}}{2},}  \label{eigenvalues sympletic}
\end{equation}%
where $\Delta (\boldsymbol{\sigma })\equiv I_{1}+I_{2}+2I_{3}$.
In this way, the inequality (\ref{Heisenberg principle}) re-writes as: 
\begin{equation}
d_{-}\geq 1/2\text{.}  \label{minimum symplectic}
\end{equation}%
A real symmetric definite positive matrix satisfying $d_{-}\geq 1/2$ 
corresponds to a proper CM i.e. describes a physical state.
\subsection{Purity and entropies}
The purity of the two-mode Gaussian state
$\varrho$, may be expressed as a function of the CM (\ref{CM}) as follows \cite{sal0}: 
\begin{equation}
\mu \equiv \mu = \mathrm{Tr}[\varrho^2] =\left(16\, I_4\right)^{-\frac12}.
\end{equation}
Another quantity, characterizing the degree of mixedness of $\varrho$, is
the von Neumann entropy
$S\left( \varrho \right) =-\text{Tr}\left( \varrho \log \varrho
\right)$. 
If the state is pure the entropy is zero $\left( S=0\right) $, otherwise
it is positive $\left( S>0\right)$ and for two-mode Gaussian states it
may be written as \cite{ser:04,AOP05}: 
$S(\varrho)\equiv S(\boldsymbol{\sigma })=f(d_{+})+f(d_{-})$
where the symplectic eigenvalues $d_{\pm }$ are given in 
(\ref{eigenvalues sympletic}) the function 
$f(x)=(x+1/2)\log (x+1/2)-(x-1/2)\log (x-1/2)$.
It is useful to recall that for a \emph{single mode} Gaussian state the von
Neumann entropy is a function of the purity alone \cite{aga:71}: 
\begin{align}
S(\varrho) = \frac{1-\mu}{2\mu} \log\left( \frac{1+\mu}{1-\mu} \right)
-\log\left( \frac{2\mu}{1+\mu} \right),
\end{align}
whereas for a two-mode state all the four symplectic invariants are
involved.
\par
For a two-mode state $\varrho $ it is of interest to assess how much
information about $\varrho $ one can obtain by addressing the single
parties. This is of course related to the correlation between the two 
modes and can be quantified by means of the quantum mutual information 
or the conditional entropies \cite{slep:71}.
Given a two-mode state $\varrho$ the quantum mutual information
$I(\varrho)$ is defined starting from the von Neumann entropies as: 
\begin{equation*}
I\left( \varrho \right) =S\left( \varrho _{1}\right) +
S\left( \varrho _{2}\right) -S\left( \varrho \right) ,
\end{equation*}%
where $\varrho _{k}=\mathrm{Tr}_{h}(\varrho )$, with $k,h=1,2$ and $h\neq k$,
are the partial traces, i.e. the density matrices of mode $k$, as obtained 
tracing over the other mode. $I\left( \varrho \right)$ can be easily expressed 
in terms of the blocks of ${\boldsymbol{\sigma }}$ and its 
symplectic eigenvalues. One has 
\begin{equation}
I(\bmsigma )=f\left( \sqrt{I_1}\right) +f\left( \sqrt{I_2}\right)
-f(d_{+})-f(d_{-}),  \label{mutual info}
\end{equation}%
where $f(x)$ is reported above.
The conditional entropies are defined accordingly as \cite{slep:71}: 
\begin{align}
S(1|2)& =S(\varrho )-S(\varrho _{2}),  \label{cond12} \\
S(2|1)& =S(\varrho )-S(\varrho _{1}).  \label{cond21}
\end{align}%
If $S(1|2)\geq 0$ (or $S(2|1)\geq 0$), the conditional entropy gives the
amount of information that the party $1$ ($2$) should send to the party $2$ (%
$1$) in order to allow for the full knowledge of the overall state $\varrho $.
If $S(1|2)<0$ (or $S(2|1)<0$), the party $1$ ($2$) does not need to send any
information to the other and, in addition, they gain $-S(1|2)$ or $-S(2|1)$ 
bits of entanglement, respectively.
This has  been proved for the case of discrete variable quantum systems
\cite{horo:05} and conjectured \cite{gen:08} for infinite dimensional
ones.
\subsection{Entanglement}
A two-mode quantum state $\rho $ is separable if and only if it can be
expressed in the following form: $\rho =\sum_{k}p_{k}\left( \rho
_{k}^{(a)}\otimes \rho _{k}^{(b)}\right) $, with $p_{k}>0$, $\sum_{k}p_{k}=1$
and $\rho _{k}^{(a)}$ and $\rho _{k}^{(b)}$ are single-mode density matrices
of the two modes $a$ and $b$, respectively. Viceversa if the state is not
separable, it is entangled. A general solution to the problem of separability 
for mixed state has not been found yet. For two-mode Gaussian states there 
exist necessary and sufficient conditions to assess whether a given
state is entangled or not. In particular, there are two equivalent
criteria, usually referred to as Duan criterion and Peres-Horodecki-Simon 
criterion, which found an explicit form in term of the CM elements. 
The criteria provide a test for entanglement, whereas to assess
quantitatively the entanglement content of a state one may use the logarithmic 
negativity or the negativity of the conditional entropies, as we see below.
\subsubsection{Duan criterion}
This criterion \cite{dua00} is based on the evaluation of the sum of the 
variances associated to a pair of EPR-like operators defined on the two 
different subsystems. For any separable continuous variable state, the 
total variance is bounded by twice the uncertainty
product. For entangled states this bound can be exceeded and the violation 
provides a necessary and sufficient condition for entanglement.
The criterion leads to an inequality that can be expressed in terms of
standard form CM elements: 
\begin{equation}
\beta_D =na^{2}+\frac{m}{a^{2}}-\left\vert c_{1}\right\vert
-\left\vert c_{2}\right\vert < a^{2}+\frac{1}{a^{2}},
\end{equation}%
with $a^{2}=\sqrt{\frac{n-1}{m-1}}$. A separable state will not satisfy the
above inequality. The criterion raises from the fact that for an entangled
state it is possible to gain information on one of the subsystems suitably
measuring the other one.
\subsubsection{Peres-Horodecki-Simon criterion (PHS)}
Also PHS criterion establishes a necessary and sufficient condition for
separability of bipartite Gaussian states \cite{sim00}. Given the CM ${%
\boldsymbol{\sigma }}$, the corresponding two-mode Gaussian state is not
separable iff: 
\begin{equation}
\tilde{{\boldsymbol{\sigma }}}+\frac{i}{2}\boldsymbol{\Omega } < 0,
\label{PHS:crit}
\end{equation}%
where $\boldsymbol{\Delta }=\mathrm{diag}[1,1,1,-1]$ and $\tilde{{%
\boldsymbol{\sigma }}}=\boldsymbol{\Delta }{\boldsymbol{\sigma }}\boldsymbol{%
\Delta }$ is the CM associated with the partially transposed density matrix.
Thanks to the symplectic invariants $\left\{
I_{1},~I_{2},~I_{3},~I_{4}\right\} $ the inequality (\ref{PHS:crit}) can be
written in a form that resembles the uncertainty relation (\ref{Heis rel.
sympl invariant}) \cite{AOP05}: 
\begin{equation}
I_{1}+I_{2}+2\left\vert I_{3}\right\vert > 4I_{4}+\frac{1}{4},
\end{equation}%
or, in terms of standard form CM elements, as: 
\begin{equation}
n^{2}+m^{2}+2\left\vert c_{1}c_{2}\right\vert -4\left( nm-c_{1}^{2}\right)
\left( nm-c_{2}^{2}\right) \leq \frac{1}{4},
\end{equation}%
or simply as: 
\begin{equation}
\tilde{d}_{-} < 1/2,
\end{equation}%
where :
\begin{equation}
\tilde{d}_{\pm }=\sqrt{\frac{\tilde{\Delta}(\boldsymbol{\sigma })\pm \sqrt{%
\tilde{\Delta}(\boldsymbol{\sigma })^{2}-4I_{4}}}{2},}  \label{symp:eig:PT}
\end{equation}%
are the symplectic eigenvalues of $\tilde{{\boldsymbol{\sigma }}}$ and $%
\tilde{\Delta}(\boldsymbol{\sigma })=I_{1}+I_{2}-2I_{3}.$ Therefore,
iff $\tilde{d}_{-}<1/2$ the Gaussian state under investigation is entangled.
\par
For an entangled state a quantitative measure of entanglement can be given 
on the observation that the larger is the violation $\tilde{d}_{-}<1/2$ 
the stronger the entanglement, or more properly, the stronger the 
resilience of entanglement to noise \cite{sal1,sal2,nm1,nm2}. 
The logarithmic negativity for a two-mode Gaussian state,
is given by \cite{vid02} (remind that $\tilde{d}_{-}>0$)
\begin{equation}
E(\bmsigma)=\max \left\{ 0,-\log 2\tilde{d}_{-}\right\} ,
\end{equation}%
and it is a simple increasing monotone function of the minimum symplectic
eigenvalue $\tilde{d}_{-}$ (for $\tilde{d}_{-}<1/2$): it thus represents a good
candidate for evaluating entanglement in a quantitative way.
It is worth to note that also the negativity of the conditional
entropies (\ref{cond12}) and (\ref{cond21}) is a sufficient condition for
entanglement \cite{cerf:99}.
\subsection{EPR correlations}
This way of assessing quantum correlations between two modes is named
after the analogy with the EPR correlation defined for a system undergone 
to a quantum nondemolition measurement (QND)\cite{prl99}. Let
us consider two subsystems $a$ and $b$, QND establishes, in principle, that
measurement performed on subsystem $b$, does not affect system $a$. This
criterion is equivalent to state that the conditional variance $V_{a|b}$ of
a quadrature of beam $a$, knowing beam $b$, takes a value smaller than the 
variance $a$ would have on its own. The conditional variance can be
expressed in terms of the unconditional variance $V_{a}$ of subsystem 
$a$ (\textit{i.e.} the
variance that the same quantity has in the subsystem $a$ space) and
normalized correlation $C_{ab}$ between the two \cite{man02,tre05}: 
$V_{a|b}=V_{a}\left( 1-C_{ab}^{2}\right)$ ,
an analogous relation holds for $V_{b|a}$. A bipartite state is said to
be EPR correlated if it verifies the following inequality: 
\begin{equation}
V_{a|b}V_{b|a} < 1/4,  \label{correlation EPR}
\end{equation}%
that can be rewritten in terms of standard form CM elements as follows: 
$\beta_{E}=nm\left( 1-\frac{c_{1}^{2}}{nm}\right) \left( 1-\frac{c_{2}^{2}%
}{nm}\right) < 1/4$.
If the inequality is satisfied in the system described by the CM 
the information on $a$ ($b$) extracted from a measurement on $b$ ($a$) is
sufficient for knowing its state with a precision better than the limit
given by the variance for a coherent state. In turn, EPR correlations are 
stronger than entanglement \cite{bow04,Rd1,dru90}, i.e. all EPR states are 
entangled whereas the converse is not true and there are entangled states 
violating Ineq. (\ref{correlation EPR}).
\section{Covariance matrix reconstruction}
\label{s:rec}
In this section we describe in some detail the method we have implemented 
to experimentally reconstruct the CM given in Eq.~(\ref{CM}) and, thus, 
to fully characterize a bipartite Gaussian state.
As expected, each autocorrelation block, $A$ or $B$, is retrieved by
measuring only the single-mode quadratures of the concerned mode $a$ or $b$.
Diagonal terms of $A$ correspond to the variances of $x_{a}$ and $y_{a}$ and
are directly available at the output of the homodyne detection. Off-diagonal
terms are instead obtained by measuring the two additional quadratures 
$z_{a}=\frac{1}{\sqrt{2}}\left( x_{a}+y_{a}\right)$, 
$t_{a}=\frac{1}{\sqrt{2}}
 \left(x_{a}-y_{a}\right)$,
and exploiting the relation $\sigma
_{12}=\sigma _{21}=\frac{1}{2}(\langle z_{a}^{2}\rangle -\langle
t_{a}^{2}\rangle )-\langle x_{a}\rangle \langle y_{a}\rangle $ \cite{sh05}.
The block $B$ is reconstructed in the same way from the quadratures of $b$.
Elements of block $C$ involve the products of quadrature of modes $a$ and $b$
and cannot be obtained by measuring individually the two modes. Instead they
are obtained by homodyning the auxiliary modes $c=\frac{1}{\sqrt{2}}(a+b)$, $%
d=\frac{1}{\sqrt{2}}(a-b)$, $e=\frac{1}{\sqrt{2}}(ia+b)$, and $f=\frac{1}{%
\sqrt{2}}(ia-b)$ and by making use of the following relations: 
\begin{eqnarray*}
\sigma _{13} &=&\frac{1}{2}(\langle x_{c}^{2}\rangle -\langle
x_{d}^{2}\rangle )-\langle x_{a}\rangle \langle x_{b}\rangle , \\
\sigma _{14} &=&\frac{1}{2}(\langle y_{e}^{2}\rangle -\langle
y_{f}^{2}\rangle )-\langle x_{a}\rangle \langle y_{b}\rangle , \\
\sigma _{23} &=&\frac{1}{2}(\langle x_{f}^{2}\rangle -\langle
x_{e}^{2}\rangle )-\langle y_{a}\rangle \langle x_{b}\rangle , \\
\sigma _{24} &=&\frac{1}{2}(\langle y_{c}^{2}\rangle -\langle
y_{d}^{2}\rangle )-\langle y_{a}\rangle \langle y_{b}\rangle
\end{eqnarray*}%
It is worth to note that since $\langle x_{f}^{2}\rangle =\langle
x_{b}^{2}\rangle +\langle y_{a}\rangle ^{2}-\langle x_{e}^{2}\rangle $ and $%
\langle y_{f}^{2}\rangle =\langle x_{a}^{2}\rangle +\langle y_{b}^{2}\rangle
-\langle y_{e}^{2}\rangle $, the measurement of the $f$-quadratures is not
mandatory.
\par
As we will see in the following, our experimental setup allows one to
mix the modes $a$ and $b$, say the signal and idler, thank to the 
polarization systems at the
OPO output. At the same time, the quadratures $x=x_{0}$, $y=x_{\pi /2}$, $%
z=x_{\pi /4}$ and $t=x_{-\pi /4}$ required for the entanglement measurement
and for the reconstruction of the CM can be easily and reliably
reconstructed from the pattern function tomography applied to data
collected in a $2\pi $ scan of the homodyne detector.
\section{Experimentals}
\label{s:exp}
\begin{figure}[htb]
\includegraphics[width=0.95\columnwidth]{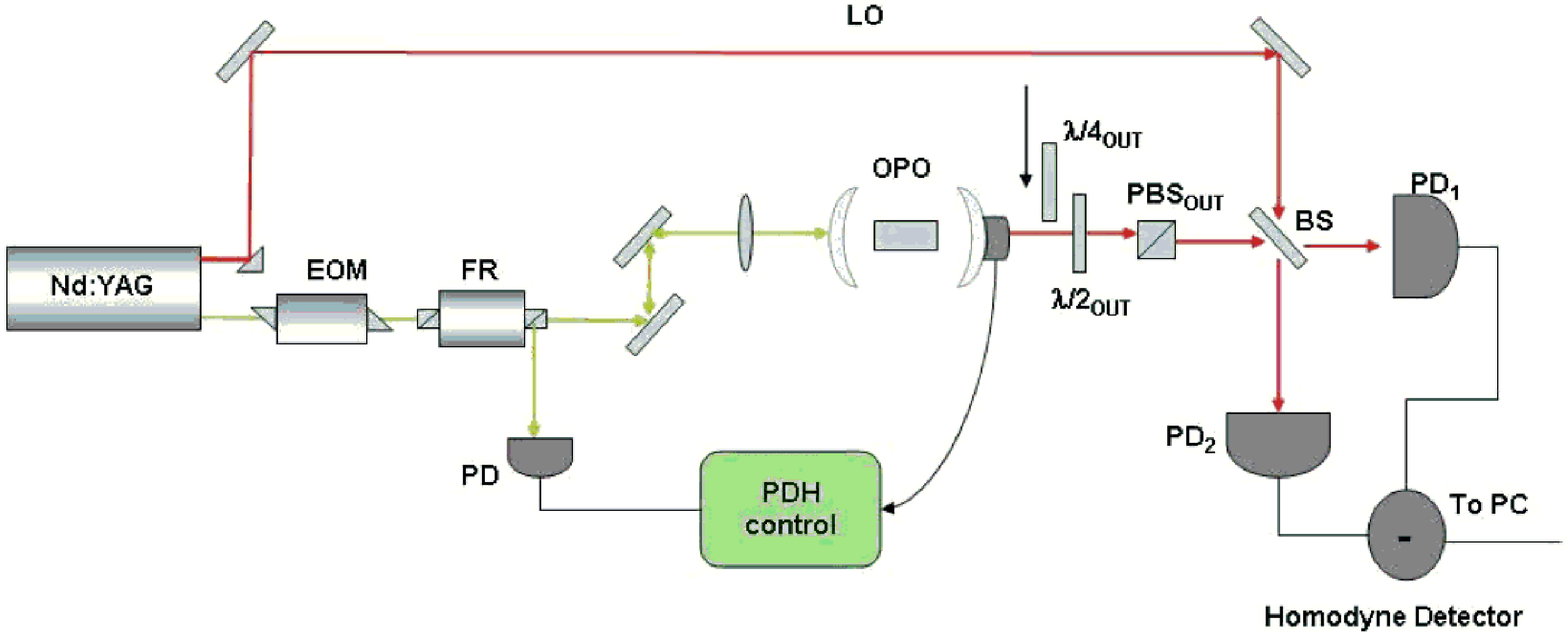}
\vspace{-0.3cm}
\caption{ Experimental setup: A type-II OPO containing a
periodically poled crystal (PPKTP) is pumped by the second harmonic of a
Nd:YAG laser. At the OPO output, a half-wave plate
($\lambda /2_{\rm out}$), a quarter-wave plate ($\lambda /4_{\rm out}$) and a
PBS$_{\rm out}$ select the mode for homodyning.
The resulting electronic signal is acquired via a PC module.} \label{f:setup}
\end{figure}

The experimental setup is schematically depicted in Fig. \ref{f:setup}. It
is based on a CW internally frequency doubled Nd:YAG
laser (Innolight Diabolo) whose outputs @532nm and @1064nm are respectively
used as the pump for a non degenerate optical parametric oscillator (OPO)
and the local oscillator (LO) for the homodyne detector. The OPO is set to
work below the oscillation threshold and it provides at its output two
entangled thermal states (the signal, $a$ and the idler $b$): aim of the
work is indeed to measure the covariance matrix of these two beams.
\par
The OPO is based on an $\alpha $-cut periodically poled KTP non linear
crystal (PPKTP, \textit{Raicol Crystals Ltd}. on custom design) which allows
for implementing a type II phase matching with frequency degenerate and cross
polarised signal and idler beams, for a crystal temperature of $\approx $ 53$%
^{\circ }$C. The transmittivity of the cavity output mirror, $T_{out}$, is
chosen in order to guarantee, together with crystal losses ($\kappa $) and
other losses mechanisms ($T_{in}$), an output coupling parameter $\eta
_{out}=T_{out}/(T_{in}+\kappa )$ @1064 nm of $\approx $ 0.73, corresponding
to an experimental linewidth of $16$ MHz @1064 nm. In order to obtain a low
oscillation threshold, OPO cavity geometry is set to warrant simultaneous
resonance on the pump, the signal and the idler: pump resonance is
guaranteed by servo-assisting the OPO cavity with a Drever Pound Hall system 
\cite{DPH}, while the resonance of other beams is induced by exploiting the
natural birefringence of the KTP to tune the optical path of each beam
inside the cavity, through a fine control of the crystal temperature and
tilt \cite{dau08}. The OPO is equipped with an handmade control system able
to stabilise the non-linear crystal temperature up to 0.1mK. Measured
oscillation threshold is $50$ mW; during the acquisition the system has been
operated below threshold at $\approx $60\% of the threshold power.
\par
The signal and idler modes are then sent to the covariance matrix
measurement set-up: this consists in a preliminary polarisation system, that
allows choosing the beam to be detected and a standard homodyne detector.
The polarisation system is made of an half-wave plate ($\lambda /2$)
followed by a polarising beam splitter (PBS); the different wave-plate
orientations allow choosing the beam to be transmitted by the PBS: the
signal ($a$), the idler ($b$) or their combinations $c$ and $d$.
The other auxiliary modes $e$ and $f$ may be obtained by
inserting before the PBS an additional quarter wave plate ($\lambda /4$) 
\cite{sh05}. Acquisition times are considerably short thank to pc-driven
mechanical actuators that allow setting the $\lambda /2$ and $\lambda /4$
positions in a fast and well calibrated manner.
\par
Once a beam is selected, it goes to a homodyne detector put downstream the
PBS. This exploits, as local oscillator, the laser output @1064 nm,
previously filtered and adjusted to match the geometrical properties of the
OPO output: a typical interferometer visibility is 0.98. The LO oscillator
phase $\theta $ is spanned thanks to a piezo-mounted mirror, linearly driven
by a ramp generator which is, in turn, adjusted to obtain a $2\pi $ variation
in $200$ ms. The homodyne photodiodes (PDs, model \emph{Epitaxx ETX300} )
have both nominal quantum efficiencies of $\approx $0.91 and each is matched
to a low-noise trans-impedence AC ($>$ few KHz) amplifier. The difference
photocurrent is eventually further amplified by a low noise high gain
amplifier (\emph{MITEQ AU 1442}).
\par
In order to avoid low frequency noise, the photocurrent is demodulated with
a sinusoidal signal of frequency $\Omega $=3 MHz and low-pass filtered ($B$=$%
300$ KHz), to be sent to a PCI acquisition board (Gage 14100) that samples
it with a frequency of $10^{6}$pts/run, with 14-bit resolution. The total
electronic noise power of the acquisition chain is 16 dBm below the shot
noise level, corresponding to the a SNR $\approx $40.
\section{Gaussianity tests}
\label{s:Gtest}
Since the covariance matrix contains the full information only for 
Gaussian states, a preliminary check on the Gaussian hypothesis is 
necessary in order to validate the entire approach. At first, in order 
to asses the Gaussianity of our data set we have evaluated the 
\textit{Kurtosis excess} (or Fisher's index) is calculated. Then, once 
Gaussianity is proved, a more sophisticate test is used to check the 
statistical quality of the collected data. In particular, the 
\textit{Shapiro--Wilk} \cite{SW65} test checks whether the collected data 
come from a truly random normal distribution, i.e. whether or not the data 
ensemble is a faithful replica of a Gaussian statistical population.
We underline the importance of Gaussianity tests, which is usually 
assumed rather than actually verified experimentally on the basis of
analysis of OPO data. 
\par
The Kurtosis is the distribution fourth order moment, and can be seen as a
sort of ``peakedness'' measurement of a random probability distribution.
Compared to the Gaussian value of $3\sigma ^{2}$ (where $\sigma $ is the
standard deviation) the Kurtosis--excess $\gamma$ is defined as
\begin{equation*}
\gamma =\frac{ \sum_{i=1}^{n}\left( x_{i}-\overline{x}%
\right) ^{4}p_{i}}{\sum_{i=1}^{n}\left( x_{i}-\overline{x%
}\right) ^{2}p_{i}}-3
\end{equation*}%
where $\overline{x}$ is the mean of the data and $p_{i}$ is the probability
of the $i$--th outcome. A $\gamma =0$ distribution is Gaussian. As a
matter of fact $\gamma $ gives an immediate check on the Gaussianity of the data
ensemble, whereas it cannot say anything about accidental (or
systematic) internal correlation between data. Overall, the use
of the Kurtosis test only may not lead to a conclusive assessment
of Gaussianity.
\par
For this purpose we adopt the \textit{Shapiro-Wilk} (SW) tests, which
is suitable to test the departure of a data sample from normality. SW
tests whether a data sample $\{x_{1},\dots,x_{n}\}$ of $n$
observations comes from a normally random distributed population. The
so-called $W_{\mathrm{SW}}$-statistic is the ratio of
two estimates of the variance of a normal distribution based on the data
sample. In formula: 
\begin{equation}
W_{\mathrm{SW}}=\frac{\left[ \sum_{h=1}^{n}a_{h}x_{(h)}\right] ^{2}}{%
\sum_{h=1}^{n}(x_{h}-\overline{x})^{2}},
\end{equation}%
where $x_{(h)}$ are the ordered sample values ($x_{(h)}$ is the $h$-th
smallest value) and $a_{h}$ are weights given by \cite{SW65}:%
\begin{equation*}
\left( a_{1},...a_{n}\right) =\frac{m^{T}V^{-1}}{\left(
m^{T}V^{-1}V^{-1}m\right) ^{\frac{1}{2}}}
\end{equation*}%
with $m^{T}$ the expected values of the order statistics of random variables
sampled from the standard normal distribution, and $V$ is the covariance
matrix of the order statistics. From a mere statistical point of view, $W_{%
\mathrm{SW}}$ is an approximation of the straightness of the normal
quantile-quantile probability plot, that is a graphical technique for
determining if two data sets come from populations with a common
distribution. Notice that $W_{\mathrm{SW}}\in \left[ 0,1\right]$. One
rejects the null hypothesis of normality within a significance interval of $%
0.05$, if $p$-$W_{\mathrm{SW}}\leq 0.05$, where the $p$-$W_{\mathrm{SW}}$ is
the $p$-value of $W_{\mathrm{SW}}$ \textit{i.e.} the probability of
obtaining a result at least as extreme as the one that was actually
observed, given that the Gaussian hypothesis is true.
\par
The two tests verify two complementary aspects. Even if the SW one is
considered a faithful Gaussianity test it can fail either for a
non-Gaussian or for non truly random distributions. Once the Gaussianity of
the data is proved, by means of the Kurtosis excess $\gamma $, the SW
test is used as a test for the randomness of the data ensemble.
\par
\begin{figure}[htb]
\begin{center}
\includegraphics[width=0.95\columnwidth]{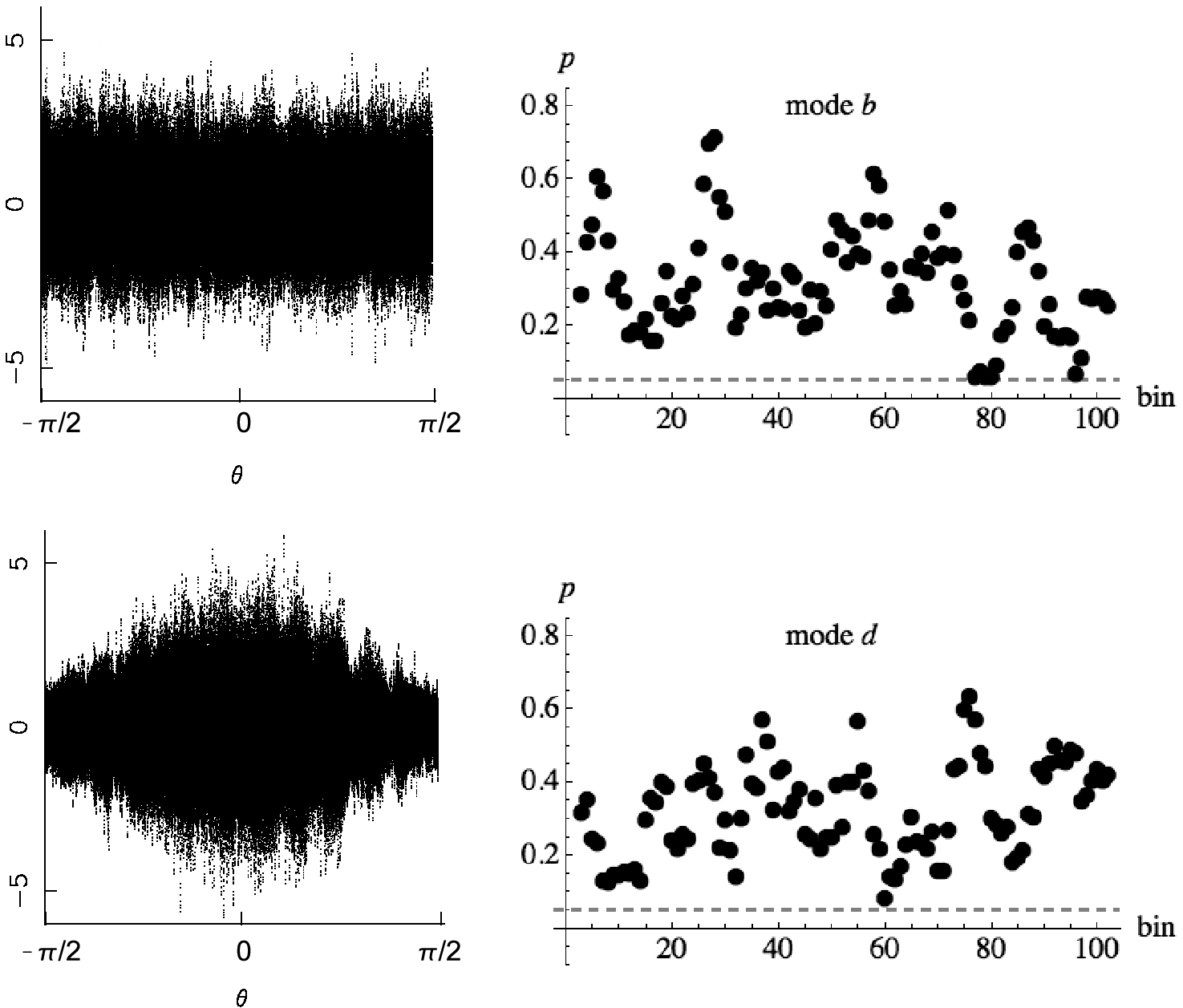}
\end{center}
\caption{(Left): from top to bottom, two typical experimental homodyne
traces of modes $b$ and $d$ (similar results are obtained for the other
modes). (Right): $p$-value of the Shapiro-Wilk normality test as a function
of the bin number (see the text for details). Since we have $p$-value $\ge
0.05$ (the dashed line in the plots), we can conclude that our data are
normally distributed. $\protect\theta$ is the relative phase between the
signal and the local oscillator. Kurtosis excess $\protect\gamma$
for these data is $0$ within experimental error.}
\label{f:tr:SW}
\end{figure}

We have applied the above statistical analysis to our homodyne data 
distribution divided into 104 discrete phase bins (each bin correspond 
to a $\theta $variation of $\approx 60~$mrad). As an example of Gaussianity
test, in Fig.~\ref{f:tr:SW} we show two typical experimental homodyne
traces for modes $b$ and $d$ (plots on the left) as well as the
corresponding $p$-value of the Shapiro-Wilk test (plots on the right).
As it is apparent from the plots, the mode $b$ is excited in a thermal
state, while the mode $c$ is squeezed with quadratures noise reduction,
corrected for non-unit efficiency, of about $2.5$~dB. An analogue
behavior has been observed for the other modes. 
Since we have $p$-value $p> 0.05$ (the dashed line in the plots) for
all the data set we can conclude that our data are normally distributed
and that the signals arriving at the detector are Gaussian states.
\par
\section{Tomographic reconstruction}
\label{s:tomo}
As already mentioned in Section \ref{s:exp} our setup is suitable to measure 
all the field quadrature $x_{\theta }=x\cos \theta +y\sin \theta $ of any 
input mode by scanning over the phase of the homodyne local oscillator.
We exploit this feature twice. On the one hand we use the full homodyne
set of data to assess Gaussianity of the state and, on the other hand,
we may perform full quantum homodyne tomography to validate results and
increase precision for some specific quantities \cite{note}.
\par
The acquisition of every mode is triggered by the PZT linear ramp: for each
value $\theta $, the quadrature $x_{\theta }=x\cos \theta +y\sin \theta $ of
the homodyne input mode is measured, where $x$\ and $y$ are respectively the
amplitude and phase field quadratures. Calibration with respect to the shot
noise is obtained by obscuring the OPO output and acquiring the vacuum
quadratures. It is worth stressing that experimentally, the acquisition over 
$2\pi $ intervals presents the advantage that it does not require
sophisticate phase locking set-up to keep $\theta $ constant during the
acquisition.
\par
The collection of homodyne data points, normalised to the shot-noise, is
then used to evaluate the bipartite state properties, included the
quadratures, $x_{\theta }$, for every $\theta $. The determination of the
quadratures mean value, as well as of any other relevant quantity, has been
performed thank to the so called \emph{pattern function} tomography. This
allows reconstructing the mean value $\langle \hat{O}\rangle $ of
an observable $\hat{O}$ as the statistical average of a suitable \emph{kernel%
} function $R[ \hat{O} ] $ over the ensemble of homodyne data $%
\left( x_{i},\theta _{i}\right)$ \cite{quo01}. By taking into account the
non-unitary detection efficiency $\eta $, $\langle \hat{O}\rangle 
$ is indeed retrieved as:%
\begin{equation*}
\langle \hat{O}\rangle =\overline{R [ \hat{O} ] }=\frac{1}{N}%
\sum_{i=1}^{N}R_{\eta }[ \hat{O}] \left( x_{i};\theta _{i}\right)
\end{equation*}%
where $N$ is the total number of samples. Every datum $\left( x_{i},\theta
_{i}\right) $ individually contributes to the average, so that the operator
mean value is gradually built up, till statistical confidence in the sampled
quantity is sufficient. Although the method is very general, and can be
applied to any operator, in the following we will only report the kernels
for the quantities we are interested in in this paper. For $\eta >0.5$, the
following kernels can be calculated (we omit the dependence of 
$R_{\eta }[ \hat{O}] (x;\theta )$ on $x$ and $\theta$)
\begin{align*}
R_{\eta }\left[ a^{\dag }a\right] &=2x^{2}-\frac{1}{2\eta }, \qquad
R_{\eta }\left[ \left( a^{\dag }a\right) ^{2}\right]  =\frac{8}{%
3}x^{4}-2x^{2} \\
R_{\eta }\left[ x_{\phi }\right]  &=2x\cos \left( \phi -\theta
\right) \\
R_{\eta }\left[ x_{\phi }^{2}\right]  &=\frac{1}{4}%
\left\{ 1+\left( 4x^{2}-\frac{1}{\eta }\right) \left[ 4\cos ^{2}\left( \phi
-\theta \right) -1\right] \right\}
\end{align*}
In principle, a precise knowledge of the $\langle \hat{O}\rangle $
would require an infinite number of measurements on equally prepared states.
However, in real experiments the number of data $N$ is of course finite, so
requiring an errors estimation. Under the hypotheses of the central limit
theorem the confidence interval on the tomographic reconstruction is given
as:%
\begin{equation*}
\delta \hat{O}=\frac{1}{\sqrt{N}}
\sqrt{\overline{\Delta R_{\eta }^{2}[ \hat{O} ] }}
\end{equation*}%
where $\overline{\Delta R_{\eta }^{2}[ \hat{O}] }$ is the kernel
variance, say the average over the tomographic data of the quantity
$R_{\eta}^{2} [ \hat{O} ] \left( x,\phi \right) -\langle \hat{O}\rangle^{2} $.
For the particular case of a field quadrature, the confidence interval is: 
\begin{equation*}
\Delta R_{\eta }\left[ x_{\theta }\right] \left( x,\phi \right)
^{2}=\left\langle \Delta x_{\theta }^{2}\right\rangle +\frac{1}{2}%
\left\langle n\right\rangle +\frac{2-\eta }{4\eta }
\end{equation*}%
where $\left\langle n\right\rangle $ is the mean photon number of the
field under scrutiny.
\section{Experimental results}
\label{s:results}
The first step is the Gaussianity test for the each data set which
consists, for each acquisition, of a collection of eight homodyne
traces: one for the shot-noise (vacuum), one for the electronic noise 
and six corresponding to the six homodyne modes $\{a,~b,~c,~d,~e,~f\}$ 
Then we check the consistency of the vacuum (shot noise) CM, namely,
$\boldsymbol{\sigma }_{0}=\frac{1}{2} {\rm Diag}(1,1)$
within the experimental errors. After this the thermal character of $a$ and $b$, 
as expected for a below threshold OPO, is verified and their mean photon number 
as well as $A$ and $B $ CM blocks are retrieved. Then, modes $c$, $d$ and 
$e$, $f$, are analysed in view of their squeezed thermal nature, with 
squeezing appearing on the $x$, $y\,$and $t$, $z$ quadratures respectively. 
The variance of $x_{c}$ (squeezed), $y_{c}$ (anti--squeezed), $x_{d}$ 
(anti--squeezed), $y_{d}$ (squeezed), $x_{e}$, $y_{e}$, $x_{f}$, and $y_{f}$, 
are finally used to retrieve the CM $C$ block.
\par
Since modes $a$ and $b$ are both, phase independent, thermal states, the
determination of their quadrature variances are highly robust against
homodyne phase fluctuations. Accordingly, the error on blocks $A$ and $B$
elements is obtained by propagating the relative tomographic error. On the
other hand, when dealing with, phase dependent, squeezed states, a small
uncertainty in setting the LO phase $\theta $ can result in a non negligible
indeterminacy on the quadrature variance used to reconstruct the relative $%
\boldsymbol{\sigma }$\ element. As a consequence, when evaluating the errors
on the elements of the block $C$, one must take into account the noise
properties of the involved modes and critically compare the tomographic
error with the error due to the finite accuracy on $\theta $. $\sigma _{13}$
and $\sigma _{24}$ are obtained as combinations of squeezed/anti--squeezed
variances, which, are stationary points of the variance as function of $%
\theta $, thus they are quite insensitive to $\theta $ fluctuations;
accordingly the overall tomographic error can be reliably used in this case.
On the contrary $\sigma _{14}$ and $\sigma _{23}$ depend on the
determination of $x_{e,f}$ and $y_{e,f}$. These quadrature variances are
extremely sensible to phase fluctuations being the variance derivative, in $%
\theta $, maximum for this values. In this case the error correspond to the
deviation between the variances at $x_{\frac{\pi }{4}}$ and at $x_{\frac{\pi 
}{4}\pm \delta \theta }$ (or $x_{-\frac{\pi }{4}}$ and $x_{-\frac{\pi }{4}%
\pm \delta \theta }$) with $\delta \theta \simeq 20$ mrad, corresponding to
the experimental phase stability of the homodyne detection.
\par
Once the full CM is reconstructed the analysis of the bipartite state can
start. First, if a failure of the uncertainty condition for the minimum
symplectic eigenvalue (see Eq. (\ref{minimum symplectic})) occurs it means
that the measurement must be discharged. In this case the reconstructed 
$\boldsymbol{\sigma }$ does not correspond to a physical state.
\par
A typical matrix is given by%
\begin{equation}\label{IV}
\boldsymbol{\sigma }=\left( 
\begin{tabular}{cccc}
1.694 & 0.000 & 1.204 & -0.02 \\ 
0.000 & 1.694 & 0.02 & -1.232 \\ 
1.204 & 0.02 & 1.671 & 0.000 \\ 
-0.02 & -1.232 & 0.000 & 1.671%
\end{tabular}%
\ \right)\:.
\end{equation}%
It corresponds to an entangled state that satisfies the Duan criterion 
($\beta_{D}=0.93$) and the Simon criterion ($\tilde{d}_{-}=0.46$,
$E(\boldsymbol{\sigma})=0.12$) while it
it does not show EPR correlations  ($\beta_{EPR}=0.65$). This fact is not
surprising: the state is rather robust (the mean photon number of the system
is $\approx $2.4) but the correlation, while ensuring the non--separability
of the state, does not provide EPR--like effect. Such a state would not be
useful in quantum protocols where EPR is exploited whereas it is 
sufficiently correlated for enhancing the security of CV QKD. For such a state 
the entropies are given by $S(\varrho)=2.23$, $ S(1|2)=0.734$ and
$S(2|1)=0.720$ and the quantum mutual information by
$I(\bmsigma)=0.779$.
\par
A strongly correlated system is the one corresponding to a diffenre data
set and described by the matrix%
\begin{equation}\label{tomo8}
{\boldsymbol{\sigma }}=\left( 
\begin{tabular}{cccc}
2.107 & 0.000 & 1.830 & -0.1 \\ 
0.000 & 2.107 & 0.08 & -1.573 \\ 
1.830 & 0.08 & 1.867 & 0.000 \\ 
-0.1 & -1.573 & 0.000 & 1.867%
\end{tabular}%
\right)
\end{equation}%
In this case the corresponding state, whose total energy is
$n_{tot}\approx 2.9$, is both entangled and EPR correlated
($\beta_{D}=0.64$, $\tilde{d}_{-}=0.23$, $E(\boldsymbol{\sigma})=1.12$, 
and $\beta_{EPR}=0.22$). This
reflects in a higher value for the mutual information
$I(\bmsigma)=1.633$ carried by t he state. Notice that this state
suffer from non--zero entries on the anti--diagonal elements of the CM.
This is due to a non--perfect alignment of the non--linear crystal that
give raise to a projection of a residual component of the field
polarized along $a$ onto the orthogonal polarization (say along $b$),
thus leading to a mixing among the modes \cite{dau08}. This effect is
the well known polarization cross-talk.
\par
Indeed, in the ideal case the OPO output is in a twin-beam state
$\mathbf{S}(\zeta )|0\rangle$, ${\mathbf S}(\zeta) = \exp\{\zeta a^\dag
b^\dag - \bar\zeta ab\}$ being the entangling two-mode squeezing
operator: the corresponding CM has diagonal blocks $A$, $B$, $C$ with
the two diagonal elements of each block equal in absolute value. In
realistic OPOs, cavity and crystal losses lead to a mixed state, {\em
i.e.} to an effective thermal contribution. In addition, spurious
nonlinear processes, not perfectly suppressed by the phase matching, may
combine to the down conversion, contributing with local squeezings.
Finally, due to small misalignments of the nonlinear crystal, a residual
component of the field polarized along $a$ may project onto the
orthogonal polarization (say along $b$), thus leading to a mixing among
the modes \cite{dau08}. Overall, the state at the output  is expected to
be a zero amplitude Gaussian entangled state, whose general form may be
written as
$\varrho = {\mathbf U} (\beta) {\mathbf S}(\zeta)\, 
{\mathbf{LS}}(\xi_1,\xi_2)\, {\mathbf T}\, 
{\mathbf{LS}}^\dag(\xi_1,\xi_2)\, {\mathbf S}^\dag (\zeta) 
{\mathbf U}^\dag (\beta)$, where
${\mathbf T}= \tau_1 \otimes \tau_2$, with 
$\tau_k = (1+ \bar n_k)^{-1} [\bar n_k/(1+\bar n_k )]^{a^\dag a}$ 
denotes a two-mode thermal state with $\bar n_k$ average photons
per mode,  ${\mathbf{LS}}(\xi_1,\xi_2)= S(\xi_1) \otimes S(\xi_2)$,
$S(\xi_k)=\exp\{\frac12 (\xi_k a^{\dag 2} - \bar\xi_k a^2)\}$ denotes
local squeezing and ${\mathbf U}(\beta)= \exp\{\beta a^\dag b 
- \bar\beta ab^\dag\}$ a mixing operator, $\zeta$, $\xi_k$ and $\beta$
being complex numbers. 
For our configuration, besides a thermal contribution due to internal
and coupling losses, we expect a relevant entangling contribution
with a small residual local squeezing and, as mentioned above, a possible 
mixing among the modes. 
\par
\begin{figure}[htb]
\centerline{\includegraphics[width=0.95\columnwidth]{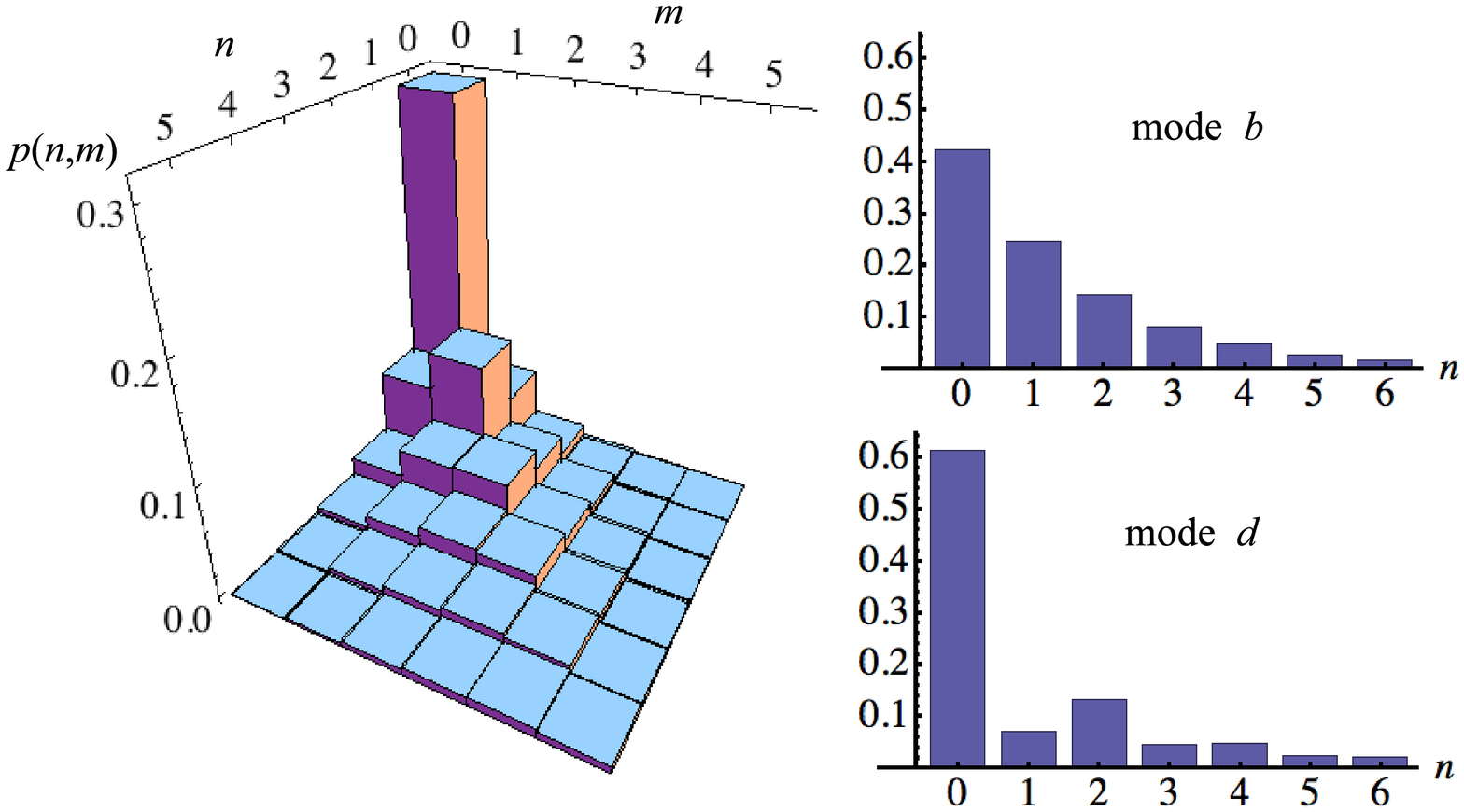}} \vspace{-0.3cm}
\caption{(Left): Joint photon number distribution $p(n,m)$ for the entangled
state of modes $a$ and $b$ at the output of the OPO. (Right): single-mode
photon distributions $p(n)$ for modes $b$ (top right) and $d$ (bottom
right). The single-mode distributions of mode $b$ is thermal and corresponds
to the marginals of $p(n,m)$. The distributions for modes $d$ is that of
squeezed thermal state. }
\label{f:pnm}
\end{figure}

Given the CM it is also possible to retrieve the corresponding joint photon
number distribution $p(n,m)$ by using the relation \cite{AOP05}: 
\begin{equation}
p(n,m) = \int_{{\mathbbm C}^2}\frac{d^2\lambda_1\,d^2\lambda_2}{\pi^2}\,
\chi(\lambda_1,\lambda_2)\,\chi_n(-\lambda_1)\,\chi_m(-\lambda_2),
\end{equation}
where $\chi(\lambda_1,\lambda_2)$ is the characteristic of the reconstructed
two-mode state, that actually depends only on ${\boldsymbol{\sigma}}$, and $%
\chi_n(\lambda_k)$ denotes the characteristic function of the projector 
$| h \rangle \langle h |$, $\chi_n(\lambda)=\langle n | 
D(\lambda) | n \rangle = \exp\{-\frac12
|\lambda|^2\} L_n(|\lambda|^2)$, 
where $L_n(x)$ is the $n$-th Laguerre polynomials.
In Fig.~\ref{f:pnm} we report the joint photon number distribution
$p(n,m)$ derived from the CM (\ref{tomo8}) and the single-mode photon
distributions (either from data or from the single-mode CM) for modes
$b$ and $d$ (the same modes considered in Fig.~\ref{f:tr:SW}): as one
may expect, the photon number distribution $b$ is thermal, whereas the
statistics of mode $d$ correctly reproduces the even-odd oscillations
expected for squeezed thermal states.
\par
The mutual information $I(\bmsigma)$ of Eq.~(\ref{mutual info}) measures the
amount of information one can get on one of the two subsystems by measuring
the other one. In turn, it is a measure of the degree of correlation
between the two modes. On the other hand, equally entangled states may show 
different $I(\bmsigma)$ and the difference appears to be dependent on the total 
number of photons. In Tab.~\ref{t:mutual_vs_ntot} we report the mutual
information $I(\bmsigma)$, the total number of photons $n_{tot}$, the Duan
and EPR factors $\beta_D$ and $\beta_E$, and the
symplectic eigenvalue $\tilde{d}_{-}$ for different acquisitions. All
the states are non--separable and not EPR-correlated; they have different 
number of photons and, correspondingly, different quantum mutual 
information.

\begin{table}[htb] \centering%
\begin{tabular}{c|c|c|c|c|}
$I(\bmsigma)$ & $n_{tot}$ & $\beta _{D}$ & $\tilde{d}_{-}$ & $\beta_E$\\ \hline\hline
0.821 & 1.421 & 1.54 & 0.34 & 0.33 \\ \hline
0.480 & 1.025 & 1.71 & 0.42 & 0.44 \\ \hline
0.366 & 0.879 & 1.84 & 0.45 & 0.46 \\ \hline
0.338 & 0.562 & 1.67 & 0.40 & 0.35 \\ \hline
0.220 & 0.485 & 1.86 & 0.44 & 0.39 \\ \hline
\end{tabular}%
\caption{The quantum mutual information for acquisition with different
average photon numbers together with the Duan and EPR factors $\beta_D$ 
and $\beta_E$, and the symplectic eigenvalue $\tilde{d}_{-}$.  
All the states are non--separable and not EPR-correlated. They have different 
number of photons and, correspondingly, different quantum mutual 
information. $I(\bmsigma)$ appears to be an increasing function of the
total energy of the state. \label{t:mutual_vs_ntot}}%

\end{table}%

\section{Conclusions}
\label{s:concl}
Gaussian states of bipartite continuous variable optical systems 
are basic tools to implement quantum information protocols and their
complete characterization, obtained by reconstructing the corresponding
covariance matrix, is a pillar for the development of quantum technology.
As a matter of fact, much theoretical attention have been devoted to
continuous variable systems and to the characterization of Gaussian
states via the CM. On the other hand, only a few experimental
reconstructions of CM have been so far reported due to the difficulties
connected to this measurement.
\par
We have developed and demonstrated a reliable and robust approach, based
on the use of a single homodyne detector, which have been tested on the
bipartite states at the output of a sub-threshold type--II OPO producing
thermal cross--polarized entangled CW frequency
degenerate beams.  The method provides a reliable reconstruction of the
covariance matrix and allows one to retrieve all the physical information
about the state under investigation. These include observable
quantities, as energy and squeezing, as well as non observable ones as
purity, entropy and entanglement.  Our procedure also includes advanced
tests for the Gaussianity of the state and, overall, represents a powerful
tool to study bipartite Gaussian states from the generation stage to the
detection one.
\section*{Acknowledgments}
The authors thank S.~Solimeno for encouragement and support. SO and 
MGAP thank M.~G.~Genoni for useful discussions. This work has been partially 
supported by the CNR-CNISM agreement.

\if \ListCaptions y

\pagebreak

{\bf List of captions}

\begin{itemize}

\item Fig.~\ref{f:setup}.
Experimental setup: A type-II OPO containing a
periodically poled crystal (PPKTP) is pumped by the second harmonic of a
Nd:YAG laser. At the OPO output, a half-wave plate
($\lambda /2_{\rm out}$), a quarter-wave plate ($\lambda /4_{\rm out}$) and a
PBS$_{\rm out}$ select the mode for homodyning.
The resulting electronic signal is acquired via a PC module.

\item Fig.~\ref{f:tr:SW}.
(Left): from top to bottom, two typical experimental homodyne
traces of modes $b$ and $d$ (similar results are obtained for the other
modes). (Right): $p$-value of the Shapiro-Wilk normality test as a function
of the bin number (see the text for details). Since we have $p$-value $\ge
0.05$ (the dashed line in the plots), we can conclude that our data are
normally distributed. $\protect\theta$ is the relative phase between the
signal and the local oscillator. Kurtosis excess $\protect\gamma$
for these data is $0$ within experimental error.

\item Fig.~\ref{f:pnm}.
(Left): Joint photon number distribution $p(n,m)$ for the entangled
state of modes $a$ and $b$ at the output of the OPO. (Right): single-mode
photon distributions $p(n)$ for modes $b$ (top right) and $d$ (bottom
right). The single-mode distributions of mode $b$ is thermal and corresponds
to the marginals of $p(n,m)$. The distributions for modes $d$ is that of
squeezed thermal state.

\end{itemize}

\pagebreak

\fi

\end{document}